\documentclass[aip, preprint,showpacs,showkeys]{revtex4-1}
\usepackage{amssymb, dcolumn,  graphicx, latexsym, amsfonts, bm}

\begin{document}

\title{The temporal evolution of the kinetic drift-Alfven instability of plasma shear flow}
\author{V. V. Mikhailenko}\email[E-mail: ]{vladimir@pusan.ac.kr}
\affiliation{Pusan National University,  Busan 609--735, S. Korea.}
\author{V. S. Mikhailenko}
\affiliation{V.N. Karazin Kharkov National University, 61108 Kharkov, Ukraine.}
\affiliation{Kharkiv National Automobile and Highway University, 61002 Kharkov, Ukraine.}
\author{Hae June Lee}
\affiliation{Pusan National University, Busan 609--735, S. Korea.}
\author{M.E.Koepke}
\affiliation{Department of Physics, West Virginia University, Morgantown, WV, USA}


\begin{abstract}
The linear non-modal kinetic theory of the kinetic drift-Alfven instability of plasma shear flows 
reveals the temporal non-modal growth with growing with time growth rate. The turbulent scattering of the sheared modes on ions accelerates this growth. The instability ceases its growth when the coupling of the drift and Alfven waves violates due to the changing with time frequencies of the drift and Alfven waves in shear flow.
\end{abstract}

\maketitle


\section{INTRODUCTION}

The electrostatic drift wave possesses the unique feature of being
unstable both in the fluid domain of a resistive plasma and in
the collision-less kinetic description. In both cases the electrons
play an essential role; in the former case, the instability is due to
their collisions with heavier species, while in the latter case the
instability arises due to the Cherenkov type electron interaction
with the parallel wave-phase velocity.
For plasmas with a not so small plasma β (>me /mi ), the
mode becomes electromagnetic and in such plasmas an inter-
play between the electrostatic and electromagnetic part of the
mode takes place. The electromagnetic component is an Alfvén-
type perturbation which is dispersive like the kinetic-Alfvén
wave (KAW) or the inertial Alfvén wave (IAW). Which of these
modes enter the perturbations depends essentially on the plasma
β, yet it also implies a certain range of the wave numbers k⊥
i

Shear flows at the plasma edge layers are of critical importance in the development of the 
regimes of high confinement mode (H-mode). The H-mode is defined by a sudden suppression 
of edge turbulence which results in an increase of the edge gradients indicating the set up of a transport barrier\cite{Wagner}. Normally H-mode is accompanied by the appearance of a quasi-periodic series of relaxation oscillations involving bursts of MHD activity known as edge localized modes (ELMs)\cite{Zohm}, development of which resulted in a periodic destruction of transport barrier and loss of considerable amount of particles and energy from 
the edge of plasma. 

The tokamak edge turbulence is at most weakly collisional. Furthermore, the plasma beta is of the order of the electron to ion mass ratio thus introducing electromagnetic effects\cite{Scott}. Therefore, it is to be expected that weakly collisional drift- Alfven turbulence i.e., drift wave turbulence coupled to kinetic Alfven waves, plays an important role in the tokamak edge. Obviously, this kind of plasma dynamics should be treated both electromagnetically and kinetically\cite {Jenko}. Despite it is known that poloidal sheared flows suppress the electrostatic drift instabilities, the effects of the sheared flows on the small scale electromagnetic instabilities is still poorly understood. It was obtained in Ref.\cite{Mikhailenko2002}, that contrary to the electrostatic drift instabilities, the drift-Alfven instabilities in shear flow exhibit the non-modal temporal growth. By using the shearing modes (or the so-called non-modal) approach to the linearised set of the two-fluid MHD equations, obtained in drift approximation for the finite beta plasma, it was predicted that in plasma shear flows with hot ions, $T_{i}\lesssim T_{e}$, the hydrodynamic and resistive drift-Alfven instabilities\cite{Mikhailovsky}  display  at time $t>\left( V'_{0}\right) ^{-1}$ the non-modal growth as  $\exp \left( \gamma_{0}V'_{0}t^{2}\right)$ of the amplitudes of the electrostatic potential $\varphi$ and of the along the magnetic field component $A_{z}$ of the vector potential, where $\gamma_{0}$ is the growth rate of these instabilities in shearless plasma. The developed theory reveals that this growth is temporal and ceases at time $t>\left(V'_{0}k_{y}\rho_{s}\right)^{-1}$. 

This paper is devoted to the analytical investigations of the processes which are responsible for the temporal evolution of the kinetic drift-Alfven instability\cite{Mikhailovsky} in plasma flows with $E\times B$ velocity shear. This instability is the electromagnetic counterpart of the electrostatic kinetic drift instability of plasma shear flows, the non-modal kinetic theory for which, grounded on the method of shearing modes, was developed in Ref.\cite{Mikhailenko-2011}. The basic equation in Ref.\cite{Mikhailenko-2011} was the integral equation for the perturbed electrostatic potential, which describes the  multi-time-scales temporal evolution of the instability. The solution\cite{Mikhailenko-2011} of this integral equation reveals at time $t\gtrsim \left(V'_{0}\right)^{-1}$ the non-modal decrease with time the frequency and growth rate of the kinetic drift instability and an ultimate suppression of that instability. In the present paper, we extend the shearing modes approach in kinetic theory\cite{Mikhailenko-2011} to the kinetic theory of the Alfven  instabilities of inhomogeneous plasma shear flows. This theory uses the transformation of the Vlasov-Amper system for potentials $\varphi$ and $A_{z}$ to sheared (in spatial and velocity) coordinates convected with shear flow and accounts for by this mean the effect of spatial time-dependent distortion of plasma disturbances by shear flows. The governing equations of this kinetic theory are derived in Sec.II. They compose the system of coupled integral equations, in which velocity shear reveals as a time-dependent effect of the finite Larmor radius. 
In Sec.III, we applied these equations to the investigation of the temporal evolution of the kinetic drift-Alfven instability of plasma shear flow. We obtain that contrary to the electrostatic drift kinetic instability, the considered drift-Alfven instability experiences the non-modal transient growth with time-dependent growth rate.
The renormalized nonlinear theory of that instability, which accounted for the turbulent scattering of ions by the ensemble of the sheared drift-Alfven perturbations is considered in Section IV. A summary of the work is given in Conclusions, Section V.

\section{BASIC EQUATIONS AND TRANSFORMATIONS}
We use the Vlasov equation for species $\alpha$ ($\alpha=i$ for ions 
and $\alpha=e$ for electrons), 
immersed in crossed  spatially inhomogeneous electric field,
$\mathbf{E}_{0}\left( \hat{\mathbf{r}}\right)$
and homogeneous magnetic field $\mathbf{B}\parallel\mathbf{e}_{z}$,
\begin{eqnarray}
&\displaystyle \frac{\partial F_{\alpha}}{\partial
t}+\hat{\mathbf{v}}\frac{\partial F_{\alpha}}
{\partial\hat{\mathbf{r}}}+\frac{e}{m_{\alpha}}\left(\mathbf{E}_{0}
\left(\hat{\mathbf{r}}\right)+\mathbf{E}_{1}
\left(\hat{\mathbf{r}}, t\right)
+\frac{1}{c}\left[\hat{\mathbf{v}}\times\left( \mathbf{B_{0}}+\mathbf{B}_{1}
\left(\hat{\mathbf{r}}, t\right)\right) \right]
\right)\frac{\partial
F_{\alpha}}{\partial\hat{\mathbf{v}}}=0\label{1}
\end{eqnarray}
in a slab geometry with the mapping $\left(r,\theta,
\varphi\right)\rightarrow \left(\hat{x}, \hat{y}, \hat{z}\right)$ where $r,\theta,
\varphi$ are the radial, poloidal and toroidal directions,
respectively, of the toroidal coordinate system. We consider the 
case of plasma shear flow in linearly changing electric field $\mathbf{E}_{0}
\left( \hat{\mathbf{r}}\right)= \left( \partial
E_{0}/\partial \hat{x}\right) \hat{x}\mathbf{e}_{x}$ with  $ \partial
E_{0}/\partial \hat{x}=const$ . In that case
\begin{eqnarray}
&\displaystyle \mathbf{V}_{0}\left(\mathbf{r}
\right)=V_{0}\left(\hat{x}\right)\mathbf{e}_{y} =-\frac{c}{B_{0}}\frac{\partial
E_{0}}{\partial \hat{x}}\hat{x}\mathbf{e}_{y} =V'_{0}\hat{x}\mathbf{e}_{y}\label{2}
\end{eqnarray}
with spatially homogeneous, $V'_{0}=const$, velocity shear. The spatially homogeneous 
part of shear flow velocity is eliminated from the problem by a simple Galilean transformation.
The fluctuating electric field $\mathbf{E}_{1}=-\nabla \varphi-\left(1/c\right)
\partial \mathbf{A}_{z}/\partial t$, and the fluctuating magnetic field, $\mathbf{B}_{1}
=\nabla\times\mathbf{A}_{z}$ in Eq.(\ref{1}) are determined with Poisson equation for the electrostatic potential $\varphi$,
\begin{eqnarray}
&\displaystyle \vartriangle \varphi\left(\textbf{r},t\right)=
-4\pi\sum_{\alpha=i,e} e_{\alpha}\int f_{\alpha}\left(\textbf{v},
\textbf{r}, t \right)d\textbf {v}_{\alpha}, \label{3}
\end{eqnarray}
and Amper law for the along magnetic field $\mathbf{B}_{0}$ component $A_{z}$ of the vector potential $\mathbf{A}$,
\begin{eqnarray}
&\displaystyle \vartriangle A_{z}\left(\textbf{r},t\right)=
-\frac{4\pi}{c}\sum_{\alpha=i,e} e_{\alpha}\int v_{z}f_{\alpha}\left(\textbf{v},
\textbf{r}, t \right)d\textbf {v}_{\alpha}, \label{4}
\end{eqnarray}
where $f_{\alpha}$ is the fluctuating part of the distribution function $F_{\alpha}$ ($f_{\alpha}=F_{\alpha}-F_{0\alpha}$, where $F_{0\alpha}$ is the ensemble average part of $F_{\alpha}$).

It was obtained in Ref.\cite{Mikhailenko-2011}, that the transition in 
the Vlasov equation from velocity $\hat{\mathbf{v}}$  and coordinates $\hat{x}$,~$\hat{y}$,~$\hat{z}$
to convected coordinates $\mathbf{v}$ in velocity space, determined by 
\begin{eqnarray}
&\displaystyle \hat{v}_{x}=v_{x},\,\,  \hat{v}_{y}=v_{y}+V'_{0}x,\,\,  \hat{v}_{z}=v_{z}
\label{5}
\end{eqnarray}
and to sheared by flow coordinates $x$,~$y$,~$z$ in the configurational space, determined by
\begin{eqnarray}
&\displaystyle 
\hat{x}=x, \,\,\,\hat{y}=y+V'_{0}tx,\,\,\hat{z}=z, 
\label{6}
\end{eqnarray}
(it is assumed that inhomogeneous electric field, and respectively shear flow, originate at time $t=t_{\left(0\right)}=0$) transforms the linearized Vlasov 
equation for $f_{\alpha}=F_{\alpha}-F_{0\alpha}$, with 
known equilibrium distribution $F_{0\alpha}$, to the form, which is free from the 
spatial inhomogeneities originated from shear flow (see also Eq.(8) in Ref.\cite{Mikhailenko-2011}). 
In the case of the electromagnetic perturbations considered here, that equation is
\begin{eqnarray}
&\displaystyle \frac{\partial f_{\alpha}}{\partial t}+v_{\alpha
x}\frac{\partial f_{\alpha}}{\partial x} +\left(v_{\alpha
y}-v_{\alpha x}V'_{0}t \right) \frac{\partial
f_{\alpha}}{\partial y}+ v_{\alpha
z}\frac{\partial f_{\alpha}}{\partial z_{\alpha}} + \omega_{c\alpha} v_{\alpha y}
\frac{\partial f_{\alpha}}{\partial v_{\alpha
x}}-\left(\omega_{c\alpha}+V'_{0} \right) v_{\alpha
x}\frac{\partial f_{\alpha}}{\partial v_{\alpha y}} \nonumber
\\  &\displaystyle
=\frac{e_{\alpha}}{m_{\alpha}}\left(\frac{\partial \varphi}{\partial
x} -V'_{0}t\frac{\partial \varphi}{\partial y} \right)
\frac{\partial F_{0\alpha}}{\partial v_{\alpha x}}
+\frac{e_{\alpha}}{m_{\alpha}} \frac{\partial \varphi}{\partial y}
\frac{\partial F_{0\alpha}}{\partial v_{\alpha y}}
+\frac{e_{\alpha}}{m_{\alpha}} \frac{\partial \varphi}{\partial
z_{\alpha}} \frac{\partial F_{0\alpha}}{\partial v_{\alpha
z}}\nonumber
\\  &\displaystyle
+\frac{1}{c}\frac{e_{\alpha}}{m_{\alpha}}\left[\frac{\partial A_{z}}{\partial t}+ v_{\alpha x}
\left(\frac{\partial A_{z}}{\partial
x} -V'_{0}t\frac{\partial A_{z}}{\partial y} \right)+v_{\alpha y}\frac{\partial A_{z}}{\partial y} \right]\frac{\partial F_{0\alpha}}{\partial v_{\alpha
z}}\nonumber
\\  &\displaystyle -\frac{e_{\alpha}}{m_{\alpha}}\frac{v_{\alpha
z}}{c}\left(\frac{\partial A_{z}}{\partial
x} - V'_{0}t\frac{\partial A_{z}}{\partial y} \right)\frac{\partial F_{0\alpha}}{\partial v_{\alpha
x}}-\frac{e_{\alpha}}{m_{\alpha}}\frac{v_{\alpha
z}}{c}\frac{\partial A_{z}}{\partial y}\frac{\partial F_{0\alpha}}{\partial v_{\alpha
y}}.\label{7}
\end{eqnarray}
($\omega_{c}$ is the cyclotron frequency of ion (electron)). 
Now, with the  electrostatic potential $\varphi\left(\textbf{r} ,t\right) $  and vector potential $A_{z}\left(\textbf{r} ,t\right)$, determined by 
Fourier transforms over the coordinates  $x$, $y$, $z$ as 
\begin{eqnarray}
&\displaystyle \varphi \left(x, y, z,
t\right) = \int\varphi \left(k_{x}, k_{y}, k_{z} ,t
\right)e^{ik_{x}x
+ik_{y}y+ik_{z}z} dk_{x}dk_{y}dk_{z},\label{8}
\end{eqnarray}
we can obtain the equation for the separate spatial Fourier harmonic of $f_{\alpha}$. With coordinates $\hat{x}$,~$\hat{y}$,~$\hat{z}$ of the laboratory set of references, 
transformation (\ref{8}) has a form
\begin{eqnarray}
&\displaystyle \varphi \left(\hat{x},\hat{y}, \hat{z},
t\right) = \int\varphi \left(k_{x}, k_{y}, k_{z} ,t
\right)e^{ik_{x}\hat{x}
+ik_{y}\left(\hat{y}-V'_{0}t\hat{x}\right) +ik_{z}\hat{z}} dk_{x}dk_{y}dk_{z}\nonumber
\\  &\displaystyle =
\int\varphi \left(k_{x}, k_{y}, k_{z} ,t
\right)e^{i\left( k_{x}-V'_{0}tk_{y}\right)\hat{x}
+ik_{y}\hat{y} +ik_{z}\hat{z}} dk_{x}dk_{y}dk_{z}.
\label{9}
\end{eqnarray}
It follows from Eq.(\ref{9}), that separate spatial Fourier mode in convected-sheared coordinates 
becomes a sheared mode with time dependent wave number $k_{x}-V'_{0}tk_{y}$ in the laboratory frame.

It follows from Eq.(\ref{7}), that the transformation of the Vlasov equation 
to convected-sheared coordinates (\ref{5}), (\ref{6}), in which separate spatial Fourier 
harmonic may be considered, converts the spatial inhomogeneity 
into the time inhomogeneity. That equation reveals (see, also, Ref.\cite{Mikhailenko-2011}), that 
the ordinary modal solution to Eq.(\ref{7}), for which ordinary dispersion equation may 
be obtained, exists only at time $t\leqslant\left(V'_{0} \right) ^{-1}$. For larger time, 
the initial value problem, which includes also other time scales, $\left(V'_{0}\right)^{-1}$ and $\left(V'_{0}k_{y}\rho_{i}\right)^{-1}$, have to be solved. The solution of such initial value 
problem for the kinetic drift instability reveals, that electrostatic potential  becomes non-modal\cite{Mikhailenko-2011} with time dependent frequency and growth rate at time $t>\left(V'_{0}\right)^{-1}$.

With leading center coordinates $X, Y$, determined through the shearing coordinates by the relations
\begin{eqnarray}
&\displaystyle
x=X-\frac{v_{\bot}}{\sqrt{\eta}\omega_{c}}\sin
\phi, \qquad
y=Y+\frac{v_{\bot}}{\eta\omega_{c}}\cos
\phi +V'_{0}t\; \left(X-x\right), \nonumber \\
&\displaystyle
z_{1}=z-v_{z}t, \label{10}
\end{eqnarray}
where $\eta=1+V'_{0}/\omega_{c}$, and with velocity coordinates $v_{\bot}$, $\phi$,
\begin{eqnarray}
&\displaystyle v_{x}=v_{\bot}\cos\phi, \qquad v_{y}=\sqrt{\eta}v_{\bot}\sin\phi,
\label{11}
\end{eqnarray}
determined through the velocities $v_{x}$, $v_{y}$ in convective frame, the Vlasov equation (\ref{1}), 
in which species index is suppressed, transforms into the form  
\begin{eqnarray}
&\displaystyle \frac{\partial F}{\partial
t}+\frac{e}{m\eta\omega_{c}}
\left(\frac{\partial\varphi}{\partial X} \frac{\partial
F} {\partial Y}-\frac{\partial\varphi}{\partial
Y} \frac{\partial F} {\partial X}\right)\nonumber \\
&\displaystyle
+\frac{e}{m}\frac{\sqrt{\eta}\omega_{c}}{v_{\perp}}
\left(\frac{\partial\varphi}{\partial \phi_{1}} \frac{\partial
F} {\partial v_{\perp}}-\frac{\partial\varphi}{\partial
v_{\perp}}\frac{\partial F} {\partial \phi_{1}}\right)
\nonumber \\
&\displaystyle -\frac{e}{m\eta\omega_{c}}\frac{v_{z}}{c}
\left(\frac{\partial A_{z}}{\partial X} \frac{\partial
F} {\partial Y}-\frac{\partial A_{z}}{\partial
Y} \frac{\partial F} {\partial X}\right)\nonumber \\
&\displaystyle
-\frac{e}{m}\frac{\sqrt{\eta}\omega_{c}}{v_{\perp}}\frac{v_{z}}{c}
\left(\frac{\partial A_{z}}{\partial \phi_{1}} \frac{\partial
F} {\partial v_{\perp}}-\frac{\partial A_{z}}{\partial
v_{\perp}}\frac{\partial F} {\partial \phi_{1}}\right)\nonumber \\
&\displaystyle-\frac{e}{m}\left( \frac{\partial\varphi}{\partial
z_{1}} +\frac{1}{c}\frac{\partial A_{z}}{\partial t}+\cos \phi\frac{v_{\perp}}{c}
\left(\frac{\partial A_{z}}{\partial X} -V'_{0}t\frac{\partial A_{z}}{\partial Y}\right)\right.\nonumber \\
&\displaystyle \left.+\sqrt{\eta}\frac{v_{\perp}}{c}\sin \phi \frac{\partial A_{z}}{\partial Y}\right)\frac{\partial F} {\partial v_{z}}=0,
\label{12}
\end{eqnarray}
in which any time dependent coefficients are absent. 

It follows from Eqs.(\ref{10}), (\ref{11}), that 
in shearing-convecting coordinates a particle gyro-motion is different from the gyro-motion determined in 
convective coordinates (\ref{11}) only  with spatial laboratory coordinates $\hat{x}$, $\hat{y}$.
Now it consists in the rotation with modified gyro-frequency and 
stretching of gyro-orbit along coordinate $y$ with velocity $-V'_{0}\left( x-X\right) $, 
which is negative for $x>X$ and is positive for $x<X$. With leading center coordinates (\ref{10})
the Fourier transforms (\ref{8}) for the electrostatic  potential becomes
\begin{eqnarray}
&\displaystyle \varphi \left(x, y, z,
t\right) = \int\varphi \left(k_{x}, k_{y}, k_{z} ,t
\right)\exp\left[ ik_{x}\left( X-\frac{v_{\bot}}{\sqrt{\eta}\omega_{c}}\sin
\phi\right)\right. \nonumber \\
&\displaystyle \left.+ik_{y}\left( Y+\frac{v_{\bot}}{\eta\omega_{c}}\cos
\phi\right) +iV'_{0}t\frac{k_{y}v_{\bot}}{\sqrt{\eta}\omega_{c}}\sin
\phi +ik_{z}z\right]  dk_{x}dk_{y}dk_{z} \nonumber \\
&\displaystyle = \int\varphi \left(k_{x}, k_{y}, k_{z} ,t
\right)\exp\left[ ik_{x} X +ik_{y}Y+ik_{z}z\right.\nonumber \\
&\displaystyle \left.+i\frac{k_{y}v_{\bot}\cos
\phi}{\eta\omega_{c}}-i\frac{\left(k_{x}-V'_{0}tk_{y}\right) v_{\bot}\cos
\phi}{\sqrt{\eta}\omega_{c}}\right]dk_{x}dk_{y}dk_{z} = \nonumber  \\
&\displaystyle = \int\varphi \left(k_{x}, k_{y}, k_{z}, t \right)
\exp \left[ ik_{x}X_{i}+ik_{y}Y_{i}+ik_{z}z\right.  \nonumber  \\
&\displaystyle \left. -i\frac{\hat{k}_{\bot}\left(t\right)
v_{\bot}}{\sqrt{\eta}\omega_{ci}}\sin\left(
\phi_{1}-\sqrt{\eta}\omega_{ci}t -\theta
\left(t\right)\right)\right]dk_{x}dk_{y}dk_{z},
\label{13}
\end{eqnarray}
where
\begin{eqnarray}
&\displaystyle
\hat{k}^{2}_{\perp}\left(t\right)=\left(k_{x}-V'_{0}tk_{y}\right)^{2}+\frac{1}{\eta}k_{y}^{2},\label{14}
\end{eqnarray}
and $\tan \theta =k_{y}/\sqrt{\eta}(k_{x}-V'_{0}tk_{y})$ with the same  presentation for $A_{z}\left(x, y, z,
t\right)$. 
It follows from Eq.(\ref{13}) that finite Larmor radius effect of the interaction of the perturbation 
with time independent wave numbers $k_{x},\, k_{y},\, k_{z}$ with ion, Larmor 
orbit of which is observed in sheared coordinates as a spiral continuously stretched 
with time, appears identical analytically to the interaction of the perturbation with 
wave numbers $k_{x}-V'_{0}tk_{y}, \,\,k_{y}/\sqrt{\eta}, \,\,k_{z}$  with ion, which rotates on the 
elliptical orbit that is observed in the laboratory frame. The time dependence of 
the finite Larmor radius effect is the basic linear mechanism of the action of 
the velocity shear on waves and instabilities in plasma shear flow\cite{Mikhailenko-2011}.

In what follows, we consider the equilibrium distribution function $F_{i0}$ as a Maxwellian,
\begin{eqnarray}
& \displaystyle F_{0}=\frac{n_{0}\left(X
\right)}{\left(2\pi v^{2}_{T}\right)^{3/2}}
\exp\left(-\frac{v^{2}_{\bot}+v^{2}_{z}}{v^{2}_{T}} \right),
\label{15}
\end{eqnarray}
with the inhomogeneity of the density of plasma shear flow species on coordinate $X$. 
In this paper we assume, that velocity shearing rate $V'_{0}$ is much less than 
the ion cyclotron frequency $\omega_{ci}$, and put $\eta=1$. With this approximation we 
have in Eq.(\ref{12}) that
\begin{eqnarray*}
& \displaystyle
\cos \phi\frac{v_{\perp}}{c}\left(\frac{\partial A_{z}}{\partial X} -V'_{0}t\frac{\partial 
A_{z}}{\partial Y}\right)+\sin \phi \frac{v_{\perp}}{c}\frac{\partial A_{z}}{\partial Y}=-\frac{\omega_{c}}{c}\frac{\partial A_{z}}{\partial \phi},
\end{eqnarray*}
and the solution of the Vlasov equation for the
perturbation $f\left(t,X, Y, z_{1}, v_{\bot},\phi,v_{z}\right)$ of the distribution function
$F$, $f=F-F_{0}$ with known $F_{0}$ becomes
\begin{eqnarray}
& \displaystyle f=
\frac{e}{m}\int\limits^{t}_{t_{o}}\left[\frac{1}{\omega_{c}}\left( \frac{\partial\varphi}{
\partial Y}-\frac{v_{z}}{c}\frac{\partial A_{z}}{
\partial Y}\right) \frac{\partial F_{0}}{\partial X}-\frac{\omega_{c}}{v_{\bot}}
\frac{\partial\varphi}{\partial \phi_{1}} \frac{\partial
F_{0}}{\partial v_{\bot}} \right.  \nonumber  \\
&\displaystyle \left. +\left( \frac{\partial\varphi}{\partial z_{1}}+\frac{1}{c}\frac{\partial A_{z}}{\partial t}\right)\frac{\partial F_{0}}{\partial v_{z}} \right] dt'+ f\left(X, Y, v_{\perp}, v_{z}, z_{1}, t_{0} \right). \label{16}
\end{eqnarray}
Using solution (\ref{16}) for all plasma species in Poisson equation (\ref{3})
and in Amper law (\ref{4}), we obtain the following set of integral equations,
which governs the temporal evolution of the 
separate spatial Fourier harmonic of the electrostatic potential 
$\varphi\left(\mathbf{k},t\right)$ and electromagnetic potential $A_{z}\left(\mathbf{k},t\right)$ 
in plasma shear flow, which is capable of handling linear evolution of Alfvenic instabilities of plasma 
shear flows:  
\begin{eqnarray}
&\displaystyle
\hat{k}^{2}\left( t\right)\varphi\left(\mathbf{k},t\right)=
\sum_{\alpha=i,e}\frac{i}{\lambda^{2}_{D\alpha}}\sum\limits_{n =
-\infty}^{\infty}\,\,\int\limits^{t}_{t_{0}}dt_{1}I_{n}\left(\hat{k}_{\perp}\left(t\right)\hat{k}_{\perp}\left(t_{1}\right)
\rho^{2}_{\alpha}\right)
\nonumber  \\
&\displaystyle\times\exp\left(-\frac{1}{2}\rho^{2}_{\alpha}\left(\hat{k}^{2}_{\perp}
\left(t\right)+\hat{k}^{2}_{\perp}
\left(t_{1}\right)\right)
-\frac{1}{2}k^{2}_{z}v^{2}_{T\alpha}\left(t-t_{1}\right)^{2}\right.\nonumber  \\
&\displaystyle \left.-in
\omega_{c\alpha}\left(t-t_{1}\right)
-in\left(\theta\left(t\right)-\theta\left(t_{1}\right)\right)\right)\nonumber \\
&\displaystyle \times\left[ \left(k_{y}v_{d\alpha}-n\omega_{c\alpha}
+ik^{2}_{z}v^{2}_{T\alpha}\left(t-t_{1}\right)\right)\varphi\left(\mathbf{k},t_{1}\right)\right.\nonumber \\
&\displaystyle \left.+\frac{1}{c}\left(\frac{\partial A_{z}\left(\mathbf{k},t_{1}\right)}{\partial t_{1}}+ik_{y}v_{d\alpha}A_{z}\left(\mathbf{k},t_{1}\right) \right)k_{z}v^{2}_{T\alpha}\left(t-t_{1} \right)  \right] \nonumber  \\
&\displaystyle -4\pi\sum_{\alpha=i,e}e_{\alpha}\delta
n_{\alpha}\left(\mathbf{k}, t, t_{0}\right),\label{17}
\end{eqnarray}
and
\begin{eqnarray}
&\displaystyle A_{z}\left(\mathbf{k},t\right)\frac{\hat{k}^{2}\left( t\right)c^{2}}{\omega^{2}_{pe}}=
i\int\limits^{t}_{t_{0}}dt_{1}e^{-\frac{1}{2}k^{2}_{z}v^{2}_{Te}\left(t-t_{1}\right)^{2}}\nonumber \\
&\displaystyle \times
\left[\left(i\frac{\partial A_{z}\left(\mathbf{k},t_{1}\right)}{\partial t_{1}}-k_{y}v_{de}A_{z}\left(\mathbf{k},t_{1}\right)\right)\left(1-k^{2}_{z}v^{2}_{Te}\left(t-t_{1}\right)^{2}\right)\right.\nonumber \\
&\displaystyle \left.-\left(1+ik_{y}v_{de}\left(t-t_{1}\right)-k^{2}_{z}v^{2}_{Te}\left(t-t_{1}\right)^{2}\right)ck_{z}\varphi\left(\mathbf{k},t_{1}\right) \right].\label{18}
\end{eqnarray}
In Eq. (\ref{17})
\begin{eqnarray}
&\displaystyle 4\pi\sum_{\alpha=i,e}e_{\alpha}\delta
n_{\alpha}\left(\mathbf{k}, t, t_{0}\right)=8\pi^{2}\sum_{\alpha=i,e}e_{\alpha}
\int\limits_{-\infty}^{\infty}dv_{z}e^{-ik_{z}v_{z}t}\nonumber  \\
&\displaystyle \int\limits_{0}^{\infty}
dv_{\bot}v_{\bot}J_{0}\left(\frac{\hat{k}_{\perp}\left(t\right)v_{\bot}}{
\omega_{c\alpha}} \right)f_{\alpha}\left( t=t_{0}, \mathbf{k},
v_{\perp}, v_{z} \right),\label{19}
\end{eqnarray}
and $f_{\alpha}\left( t=t_{0}, \mathbf{k}, v_{\perp}, v_{z}
\right)$ is the initial, determined at $t=t_{0}$ perturbation,
assumed here as not dependent on $\phi$, of the distribution
function $F_{\alpha}$. It follows from Eq.(\ref{19}), that initial
perturbation $\varphi\left(\mathbf{k},  t=t_{0}\right)$ of the
self-consistent electrostatic potential is equal to
\begin{eqnarray}
&\displaystyle \varphi\left(\mathbf{k},  t=t_{0}\right) =
-\frac{4\pi}{k^{2}_{\perp} \left(t_{0} \right)
+k^{2}_{z}}\sum_{\alpha=i,e}e_{\alpha}\delta
n_{\alpha}\left(\mathbf{k}, t_{0}, t_{0}\right).\label{20}
\end{eqnarray}

In Ref.\cite{Mikhailenko-2011} we found as more convenient and transparent for further analysis to use 
the explicitly causal representation of the equation for the electrostatic potential with function $\Phi\left(\textbf{k},
t\right)=\varphi\left(\textbf{k},t\right)\Theta\left(t-t_{0}\right)$, where
$\Theta\left(t-t_{0}\right)$ is the unit-step Heaviside function (it is equal to zero for $t<t_{0}$ and 
equal to unity for $t\geq t_{0}$). The equation for $\Phi\left(\textbf{k}, t\right)$, which is relevant for the analysis of the low frequency electromagnetic drift type perturbations, has a form
\begin{eqnarray}
&\displaystyle \left(k^{2}_{\bot}\left(t \right)+k^{2}_{z}\right) \Phi\left(\mathbf{k}, t\right)=
-\frac{1}{\lambda^{2}_{Di}}\int\limits^{t}_{t_{0}}dt_{1}\frac{d}{dt_{1}}\left\lbrace \Phi\left(\mathbf{k}, t_{1}\right)\left[1- A_{0i}\left(t, t_{1}\right)\right]\right\rbrace \nonumber  \\
&\displaystyle +\frac{1}{\lambda^{2}_{Di}}\int\limits^{t}_{t_{0}}dt_{1}A_{0i}\left(t, t_{1}\right)k_{y}v_{di}\Phi\left(\mathbf{k}, t_{1}\right)+\nonumber  \\
&\displaystyle +\frac{i}{\lambda^{2}_{De}}\int\limits^{t}_{t_{0}}dt_{1}e^{-\frac{1}{2}k^{2}_{z}v^{2}_{Te}\left(t-t_{1} \right)^{2}}\left\lbrace\left(k_{y}v_{de}+ik_{z}^{2}v^{2}_{Te}\left(t-t_{1}\right)\right) \Phi\left(\mathbf{k}, t_{1}\right) \right. \nonumber  \\
&\displaystyle \left.-\frac{i}{c}k_{z}v^{2}_{Te}\left(t-t_{1}\right)\left(i\frac{\partial A_{z}\left(\mathbf{k}, t_{1}\right)}{\partial t_{1}}-k_{y}v_{de} A_{z}\left(\mathbf{k}, t_{1}\right)\right) \right\rbrace 
+4\pi\sum_{\alpha}e_{\alpha}\delta n_{\alpha}\left(\mathbf{k}, t, t_{0}\right), 
\label{21}
\end{eqnarray}
where
\begin{eqnarray}
&\displaystyle A_{0i}\left(t, t_{1}\right)=
I_{0}\left(\hat{k}_{\perp}\left(t\right)\hat{k}_{\perp}\left(t_{1}\right)
\rho^{2}_{i}\right)
e^{-\frac{1}{2}\rho^{2}_{i}\left(\hat{k}^{2}_{\perp}\left(t\right)+\hat{k}^{2}_{\perp}
\left(t_{1}\right)\right)}.\label{22}
\end{eqnarray}
It follows from system (\ref{18}), (\ref{21}), that velocity shear meets only in the ions term 
$A_{0i}\left(t, t_{1}\right)$ of Eq.(\ref{18}) and in time dependent $k\left(t \right)$ of the 
left hand-side of Eq.(\ref{21}), that introduce the non-modal effects in the temporal evolution 
of the potentials. 

\section{KINETIC DRIFT-ALFVEN INSTABILITY IN SHEAR FLOW}

In this section, we consider the temporal evolution of the kinetic drift-Alfven instability
in plasma shear flow. Drift-Alfven instability involves the coupling of the density-gradient
driven drift waves with Alfven waves, when the phase velocities of the two waves become 
comparable. It is obvious, that for this instability it is impossible to obtain the explicit analytical 
solution to system (\ref{17}), (\ref{21}), as well as it was for more simple electrostatic
kinetic drift instability of plasma shear flow\cite{Mikhailenko-2011}. For the approximate solution 
of the system (\ref{17}), (\ref{21}), we consider the perturbations  for which the potentials 
$\Phi\left(\mathbf{k}, t\right)$ and $A\left(\mathbf{k}, t\right)$ change over time as  
\begin{eqnarray}
&\displaystyle\left|k_{z}v_{Ti}\right|< \left|d\ln\Phi\left(\mathbf{k}, t\right)/dt\right|
\sim \left|d\ln A\left(\mathbf{k}, t\right)/dt\right|<\left|k_{z}v_{Te}\right|,
\label{23}
\end{eqnarray}
that corresponds to the model of the hydrodynamic ions and almost adiabatic electrons in plasma with $\beta=4\pi nT_{e}/B^{2}$ (the ratio of thermal to magnetic pressure) in the 
range $1\gg\beta\gg m_{e}/m_{i}$. Introducing the time scale $t_{*}$ of the order of $\left|d\ln\Phi\left(\mathbf{k}, t\right)/dt\right|$ and 
$\left|d\ln A\left(\mathbf{k}, t\right)/dt\right|$, the condition (\ref{23}) will hold at 
time $t_{1}$ in Eq.(\ref{21})
\begin{eqnarray}
&\displaystyle\frac{1}{k^{2}_{z}v^{2}_{Ti}}\gg \left(t-t_{1}\right)^{2}\sim t^{2}_{*}\gg 
\frac{1}{k^{2}_{z}v^{2}_{Te}}.
\label{24}
\end{eqnarray}
For that time, we have $k^{2}_{z}v^{2}_{Te}\left(t-t_{1} \right)^{2}\gg 1 $ in electron 
terms of Eqs.(\ref{18}) and (\ref{21}), so that $\exp\left( -k_{z}^{2}v^{2}_{Te}\left(t-t_{1}\right)^{2}\right)$ is not small only for $t_{1}\simeq t$. In ion terms, we have $k^{2}_{z}v^{2}_{Ti}
\left(t-t_{1}\right)^{2}\ll 1$, that correspond to small ion Landau damping. 
Using the approximation $e^{-\frac{1}{2}k^{2}_{z}v^{2}_{Ti}\left(t-t_{1}\right)^{2}}\approx 1$ and  
using the Taylor series expansion for both potentials in electron terms with accounting for the terms of the order 
of $\left(t_{1}-t \right)^{2}$, the system (\ref{18}) and (\ref{21}) reduces to the equations
\begin{eqnarray}
&\displaystyle
\left[-\left(1+T \right)+A_{0i}\left(t\right) \right]\Phi\left(\mathbf{k},t\right)
\nonumber \\
&\displaystyle
+ i k_{y}v_{di}\int\limits^{t}_{t_{0}}dt_{1}
\Phi\left(\mathbf{k},t_{1}\right)A_{0i}\left(t, t_{1}\right)
\nonumber \\
&\displaystyle
+i\sqrt{\frac{\pi}{2}}\frac{k_{y}v_{de}}{k_{z}v_{Te}}\Phi\left(\mathbf{k},t\right)+T\sqrt{\frac{\pi}{2}}\frac{1}{k_{z}v_{Te}}\frac{\partial \Phi\left(\mathbf{k},t\right)}{\partial t}\nonumber \\
&\displaystyle +\frac{1}{k_{z}c}T\left( 1-\sqrt{\frac{\pi}{2}}\frac{1}{k_{z}v_{Te}}\frac{\partial}{\partial t}\right) \left(i\frac{\partial A\left(\mathbf{k},t\right)}{\partial t}-k_{y}v_{de} A\left(\mathbf{k},t\right) \right)=0, \label{25}
\end{eqnarray}
and
\begin{eqnarray}
&\displaystyle k^{2}_{\bot}\left(t\right)\rho^{2}_{s}A\left(\mathbf{k}, t\right)=\frac{i}{k^{2}_{z}v^{2}_{A}}\frac{\partial}{\partial t}
\left(i\frac{\partial A\left(\mathbf{k},t\right)}{\partial t}-k_{y}v_{de} A\left(\mathbf{k},t\right) \right)
\nonumber \\
&\displaystyle
-i\sqrt{\frac{\pi}{2}}\frac{1}{k^{2}_{z}v^{2}_{A}}\frac{1}{k_{z}v_{Te}}\frac{\partial^{2}}{\partial t^{2}}
\left(i\frac{\partial A\left(\mathbf{k},t\right)}{\partial t}-k_{y}v_{de} A\left(\mathbf{k},t\right) \right)
\nonumber \\
&\displaystyle
+ck_{z}\frac{k_{y}v_{de}}{k^{2}_{z}v^{2}_{A}}\Phi\left(\mathbf{k}, t\right)+i\sqrt{\frac{\pi}{2}}\frac{ck_{z}}{k^{2}_{z}v^{2}_{A}}\frac{1}{k_{z}v_{Te}}\frac{\partial^{2}\Phi\left(\mathbf{k}, t\right)}{\partial t^{2}}
\nonumber \\
&\displaystyle
-\frac{ick_{z}}{k^{2}_{z}v^{2}_{A}}\left(1-i\sqrt{\frac{\pi}{2}}\frac{k_{y}v_{de}}{k_{z}v_{Te}} \right)\frac{\partial \Phi\left(\mathbf{k}, t\right)}{\partial t}.\label{26} 
\end{eqnarray}
respectively. Here $T=T_{i}/T_{e}$, $v_{di(e)}= \left(cT_{i(e)}/eB\right)
d\ln n_{i0}/dx$ is the ion (electron) diamagnetic velocity,
and 
\begin{eqnarray}
&\displaystyle A_{0i}\left(t\right)=
I_{0}\left(\hat{k}^{2}_{\perp}\left(t\right)\rho^{2}_{i}\right)
e^{-\rho^{2}_{i}\hat{k}^{2}_{\perp}\left(t\right)}.\label{27}
\end{eqnarray} 
In this paper we consider the case of shear flow with small velocity shear $V'_{0}$, for which 
\begin{eqnarray}
&\displaystyle 
\left|k_{z}v_{A}\right|\gtrsim \left|k_{y}v_{de}\right|\gg \left|V'_{0}\right|\sim \gamma_{0}\label{28},
\end{eqnarray}
where $\gamma_{0}$ is the modal growth rate of the kinetic drift-Alfven instability of the shearless plasma. 
In that case, the non-modal processes, which are determined in Eq.(\ref{25}) by the time dependent $k_{\bot}\left(t\right)$ in $A_{0i}\left(t, t_{1}\right)$, are slow and may be determined by the approximation
\begin{eqnarray}
&\displaystyle \Phi\left(\mathbf{k}, t\right)=\Phi\left(\mathbf{k}\right)e^{-i\int\limits^{t}_{t_{0} }\omega\left(\mathbf{k}, t_{1} \right)dt_{1}}, \,\,\,  A\left(\mathbf{k}, t\right)=A\left(\mathbf{k}\right)e^{-i\int\limits^{t}_{t_{0} }\omega\left(\mathbf{k}, t_{1} \right)dt_{1}}.\label{29}
\end{eqnarray}
The second term in Eq.(\ref{25}), that contains the integral of $\Phi\left(\mathbf{k}, t\right)$ over time may be presented by the integration in parts in the form
\begin{eqnarray}
&\displaystyle  i k_{y}v_{di}\int\limits^{t}_{t_{0}}dt_{1}
\Phi\left(\mathbf{k}\right)e^{-i\int\limits^{t_{1}}_{t_{0} }\omega\left(\mathbf{k}, t_{2} \right)dt_{2}}A_{0i}\left(t, t_{1}\right)\nonumber \\
&\displaystyle =-\frac{k_{y}v_{di}}{\omega\left(\mathbf{k}, t\right)}\Phi\left(\mathbf{k}\right)A_{0i}\left(t\right)e^{-i\int\limits^{t}_{t_{0} }\omega\left(\mathbf{k}, t_{1} \right)dt_{1}}
+\frac{k_{y}v_{di}}{\omega\left(\mathbf{k}, t_{0}\right)}\Phi\left(\mathbf{k}\right)A_{0i}\left(t, t_{0}\right)\nonumber \\
&\displaystyle +k_{y}v_{di}\Phi\left(\mathbf{k}\right)\int\limits^{t}_{t_{0}}dt_{1}\frac{i}{\omega^{2}\left(\mathbf{k}, t_{1}\right)}\frac{d}{dt_{1}}\left(
e^{-i\int\limits^{t_{1}}_{t_{0} }\omega\left(\mathbf{k}, t_{2} \right)dt_{2}}\right)\frac{d}{dt_{1}}\left(A_{0i}\left(t, t_{1}\right)\right).\label{30}
\end{eqnarray}
For exponentially growing potential $\Phi\left(\mathbf{k}, t\right)$, i.e. when $Im \omega\left(\mathbf{k}, t \right)>0$ in (\ref{29}), the first term in Eq.(\ref{30}) at time $t-t_{0} >\left( Im \omega\left(\mathbf{k}, t \right)\right) ^{-1} $ is exponentially greater with respect to the second term. Also, the third term in the right-hand side of (\ref{30}) is in $\left|k_{y}v_{de}t\right|\gg 1$ times less then the first one for the times $t\gg \left( k_{y}v_{de}\right)^{-1}$. Therefore we retain in the right-hend side of Eq.(\ref{30}) only the first term. As a result, system (\ref{25}), (\ref{26}) reduces to the following system for amplitudes $\Phi\left(\mathbf{k}\right)$ and $A\left(\mathbf{k}\right)$:
\begin{eqnarray}
&\displaystyle \left[1+T-\left(1-\frac{k_{y}v_{di}}{\omega\left(\mathbf{k}, t\right)} \right)A_{0i}\left(t \right)
+iT\sqrt{\frac{\pi}{2}}\frac{\left( \omega\left(\mathbf{k}, t \right)-k_{y}v_{de}\right) }{k_{z}v_{Te}} \right]\Phi\left(\mathbf{k}\right)\nonumber \\
&\displaystyle
-T\frac{\left( \omega\left(\mathbf{k}, t\right)-k_{y}v_{de}\right)}{k_{z}c}\left(1+i\sqrt{\frac{\pi}{2}}\frac{\omega\left(\mathbf{k}, t\right)}{k_{z}v_{Te}} \right)A\left(\mathbf{k}\right)=0, \label{31}
\end{eqnarray}
\begin{eqnarray}
&\displaystyle 
-k^{2}_{\bot}\left(t\right)\rho^{2}_{s}A\left(\mathbf{k}\right)+\omega\left(\mathbf{k}, t\right)\left(\omega\left(\mathbf{k}, t\right)-k_{y}v_{de}\right)\frac{1}{k_{z}^{2}v^{2}_{A}}
\left(1+i\sqrt{\frac{\pi}{2}}\frac{\omega\left(\mathbf{k}, t\right)}{k_{z}v_{Te}} \right)
A\left(\mathbf{k}\right) \nonumber \\
&\displaystyle-\frac{ck_{z}}{k_{z}^{2}v^{2}_{A}}\left(\omega\left(\mathbf{k}, t\right)-k_{y}v_{de}\right)\left(1+i\sqrt{\frac{\pi}{2}}\frac{\omega\left(\mathbf{k}, t\right)}
{k_{z}v_{Te}} \right)\Phi\left(\mathbf{k}\right)=0,\label{32}
\end{eqnarray}
which gives the following dispersion equation for the time dependent frequency $\omega\left(\mathbf{k}, t\right)$:
\begin{eqnarray}
&\displaystyle \left(1-\frac{k_{y}v_{de}}{\omega\left(\mathbf{k}, t\right)} \right)\left(1+i\sqrt{\frac{\pi}{2}}\frac{\omega\left(\mathbf{k}, t\right)}
{k_{z}v_{Te}} \right)\nonumber \\
&\displaystyle \times\left(1-\frac{\omega^{2}\left(\mathbf{k}, t\right)}{k_{z}^{2}v^{2}_{A}}\left(1-\frac{k_{y}v_{di}}{\omega\left(\mathbf{k}, t\right)} \right)\frac{\left( 1-A_{0i}\left(t \right)\right)  }{k_{\bot}^{2}\left(t \right)\rho^{2}_{Ti}}  \right)\nonumber \\
&\displaystyle  +T^{-1}\left(1-\frac{k_{y}v_{di}}{\omega\left(\mathbf{k}, t\right)} \right)\left( 1-A_{0i}\left(t \right)\right) =0.   \label{33}
\end{eqnarray}
Eq.(\ref{33}) is valid for arbitrary values of $k^{2}_{\bot}\left(t\right)\rho_{Ti}^{2}$ provided that $\omega\left(\mathbf{k}, t\right)$ has a positive imaginary part, i.e. for the unstable drift-Alfven waves.
In long wavelength limit $k^{2}_{\bot}\left(t\right)\rho_{Ti}^{2}\ll 1$, Eq.(\ref{33}) reduces to the form  
\begin{eqnarray}
&\displaystyle \left(\omega\left(\mathbf{k}, t\right)-k_{y}v_{de}\right)\left(\omega^{2}\left(\mathbf{k}, t\right)-\omega\left(\mathbf{k}, t\right)k_{y}v_{di}-k_{z}^{2}v^{2}_{A}\right)\nonumber \\
&\displaystyle
=k_{z}^{2}v^{2}_{A}k^{2}_{\bot}\rho^{2}_{s}\left(\omega\left(\mathbf{k}, t\right)-k_{y}v_{di}\right)\left(1-i\sqrt{\frac{\pi}{2}}\frac{\omega\left(\mathbf{k}, t\right)}
{k_{z}v_{Te}} \right), \label{34}
\end{eqnarray}
where $\rho^{2}_{s}=\rho^{2}_{Ti}T^{-1}$. In the case, when the drift wave frequency $k_{y}v_{de}$ 
is much less then the Alfven frequency $k_{z}v_{A}$, Eq.(\ref {34}) determines the well known frequency 
$\omega_{1}$ and the growth rate $\gamma_{1}$ of the kinetic drift instability,
\begin{eqnarray}
&\displaystyle
\omega_{1}\left(\mathbf{k}, t\right)=k_{y}v_{de}\frac{1-k^{2}_{\bot}\left(t\right)\rho^{2}_{Ti}}{1+k^{2}_{\bot}\left(t\right)\rho^{2}_{s}}, \nonumber \\
&\displaystyle
\gamma_{1}\left(\mathbf{k}, t\right)=\sqrt{\frac{\pi}{2}}\frac{\left(k_{y}v_{de}\right)^{2}}{k_{z}v_{Te}}\frac{k^{2}_{\bot}\left(t\right)\rho^{2}_{s}\left(1-k^{2}_{\bot}\left(t\right)\rho^{2}_{Ti}\right)}{\left(1+k^{2}_{\bot}\left(t\right)\rho^{2}_{s}\right)^{3}}\left(1+\frac{T_{i}}{T_{e}}\right), 
\label{35}
\end{eqnarray}
and the frequencies $\omega_{2,3}\left(\mathbf{k}, t\right)$ and damping rate $\gamma_{2}\left(\mathbf{k}, t\right)$ of two kinetic Alfven  waves,
\begin{eqnarray}
&\displaystyle
\omega_{2,3}\left(\mathbf{k}, t\right)=\pm k_{z}v_{A}\left(1+k^{2}_{\bot}\left(t\right)\rho^{2}_{s}\right)^{1/2},\,\,\,\,  
\gamma_{2,3}\left(\mathbf{k}, t\right)=-\sqrt{\frac{\pi}{8}}\frac{k^{2}_{z}v^{2}_{A}}{k_{z}v_{Te}}\frac{k^{2}_{\bot}\left(t\right)\rho^{2}_{s}}{\left(1+k^{2}_{\bot}\left(t\right)\rho^{2}_{s}\right)^{1/2}}. 
\label{36}
\end{eqnarray}
It is obvious, that for the damped Alfven waves (\ref {36}) the approximation (\ref {30}) is not valid 
and the initial value solution of the system (\ref {25}), (\ref {26}) is necessary for the proper description 
of the damped perturbations. In the short wavelength limit $k_{\bot}\left(t\right)\rho_{s}\gg 1$, Eq.(\ref {33}) 
has solution for the unstable drift kinetic instability,
\begin{eqnarray}
&\displaystyle
\omega_{1}\left(\mathbf{k}, t\right)=\frac{k_{y}v_{de}}{\sqrt{2\pi}k_{\bot}\left(t\right)\rho_{Ti}}\left[1+\frac{T_{e}}{T_{i}}+\frac{T_{i}}{T_{e}}
\frac{k^{2}_{y}v_{de}^{2}}{k^{2}_{\bot}\left(t\right)k^{2}_{z}c^{2}\lambda^{2}_{Di}} \right]^{-1} \nonumber \\
&\displaystyle
\gamma_{1}\left(\mathbf{k}, t\right)=\frac{\left(k_{y}v_{de}\right)^{2}}{2k_{z}v_{Te}k_{\bot}\left(t\right)\rho_{i}}\left[1+\frac{T_{e}}{T_{i}}+\frac{T_{i}}{T_{e}}
\frac{k^{2}_{y}v_{de}^{2}}{k^{2}_{\bot}\left(t\right)k^{2}_{z}c^{2}\lambda^{2}_{Di}} \right]^{-2},
\label{37}
\end{eqnarray}
and two solutions (valid for $k_{z}v_{A}k_{\bot}\rho_{i}\ll \omega_{ci}$) for the frequency and 
the growth rate for two damped Alfven waves,
\begin{eqnarray}
&\displaystyle
\omega_{2,3}\left(\mathbf{k}, t\right)=\pm k_{z}v_{A}k_{\bot}\left(t\right)\rho_{i}\left(1+\frac{T_{e}}{T_{i}}\right)^{1/2},\,\,\,\,
\gamma_{2}\left(\mathbf{k}, t\right)= -\sqrt{\frac{\pi}{8}}\frac{k^{2}_{z}v^{2}_{A}k^{2}_{\bot}\left(t\right)\rho^{2}_{i}}{k_{z}v_{Te}}.
\label{38}
\end{eqnarray}
So, in both cases of the disparate frequencies of the drift and Alfven waves, Alfven waves are damped. 
In the case when the frequencies of the density-gradient driven drift waves and  Alfven waves becomes 
the almost equal, the kinetic drift-Alfven instability develops \cite{Mikhailovsky}.
In times $t\ll t_{s}=\left(V'_{0}k_{y}\rho_{i} \right)^{-1} $ the right hand side of 
(\ref {34}) is small. Omitting the right-hand side term in (\ref {34}) we have solutions
\begin{eqnarray}
& \displaystyle \omega_{1,2}\left(\mathbf{k}, t\right)=\frac{k_{y}v_{di}}{2}\pm\biggl(\frac
{k_{y}^{2}v_{di}^{2}}{4}+k_{z}^{2}v^{2}_{A}\biggr)^{1/2} \,\label{39} ,
\end{eqnarray}
which define two Alfven waves, and solution
\begin{eqnarray}
& \displaystyle
\omega_{3}=k_{y}v_{de},\label{40}
\end{eqnarray}
which defines the electron drift wave. In the vicinity of
the crossing of the solutions $\omega_{1}$ and $\omega_{3}$ the dispersion equation (\ref {36}) becomes
\begin{eqnarray}
& \displaystyle
\left( \delta\omega\left(\mathbf{k}, t\right)\right)^{2}\left(\omega_{1}-\omega_{2} \right)= k^{2}_{z}v^{2}_{A}k^{2}_{\bot}\left(t\right)\rho^{2}_{s}\left(\omega_{1}+\left|\omega_{2}\right| \right)\left(1-i\sqrt{\frac{\pi}{2}}\frac{\omega_{1}}{k_{z}v_{Te}} \right),
\label{41}
\end{eqnarray}
where $ \delta\omega\left(\mathbf{k}, t\right)$ is the frequency mismatch, determined as $\omega\left(\mathbf{k}, t\right)=\omega_{1}+\delta\omega\left(\mathbf{k}, t\right)=\omega_{2}+\delta\omega\left(\mathbf{k}, t\right)$.
It gives the growth rate of the kinetic drift--Alfven instability for the time $t$ in the interval $\left(V'_{0} \right)^{-1}\ll t\ll t_{s}$ and for $T_{i}=T_{e}$\cite{Mikhailovsky} 
\begin{eqnarray}
&\displaystyle \gamma\left(\mathbf{k}, t\right)=Im\delta\omega\left(\mathbf{k}, t\right)
=\sqrt{\frac{\pi}{12}}k_{y}v_{de}k_{\bot}\left(t\right)\rho_{Ti}\frac{v_{A}}{v_{Te}}. \label{42}
\end{eqnarray}
For time $t>\left(V'_{0} \right)^{-1} $, $k_{\bot}\approx V'_{0}tk_{y}$ and therefore
\begin{eqnarray}
&\displaystyle \gamma\left(\mathbf{k}, t\right)\approx\sqrt{\frac{\pi}{12}}k_{y}v_{de}k_{y}\rho_{Ti}\frac{v_{A}}{v_{Te}}V'_{0}t.
 \label{43}
\end{eqnarray}
The solutions for potentials $\Phi\left(\mathbf{k}, t\right)$ and $A\left(\mathbf{k}, t\right)$ will be of the form
\begin{eqnarray}
&\displaystyle \left( \Phi\left(\mathbf{k}, t\right), A\left(\mathbf{k}, t\right)\right)\sim\exp \left(-i\omega\left(\mathbf{k}\right)t-iRe\int^{t}\delta\omega\left(\mathbf{k}, t_{1}\right)dt_{1}\right.\nonumber \\
&\displaystyle \left.+\frac{1}{4}\sqrt{\frac{\pi}{3}}k_{y}v_{de}k_{y}\rho_{Ti}\frac{v_{A}}{v_{Te}}V'_{0}t^{2}\right).
\label{44}
\end{eqnarray}
At time $t>t_{s}$ the right-hand part of Eq.(\ref {34}) ceases to be small and the frequency mismatch becomes of the order of the Alfven frequency and disparity of the drift  and Alfven frequencies occurs. At these times drift and Alfven modes evolve separately. The drift-Alfven instability transforms into the drift kinetic instability, temporal evolution of which was considered in Ref.\cite{Mikhailenko-2011}.

\section{RENORMALIZED THEORY OF THE KINETIC DRIFT-ALFVEN INSTABILITY IN SHEAR FLOW}
In Ref.\cite{Mikhailenko-2011} we find, that the effect of the turbulent scattering of the ion Larmor gyration angle by the random electric field of the ensemble of the sheared perturbations with random phases rapidly suppresses the electrostatic drift turbulence. Here we explore the role of that effect on the temporal evolution of the kinetic drift-Alfven instability. It follows from Eq.(\ref {32}), that $A\sim \left(ck_{z}/\omega\left(\mathbf{k}, t\right)\right)\Phi$. Then, for the turbulent electric field $\mathbf{E}_{1}\left(\mathbf{r}, t\right)$ we obtain that
\begin{eqnarray}
&\displaystyle \frac{ck_{\perp}\Phi}{\omega A}\sim \frac{k_{\perp}\left(t\right)}{k_{z}}\gg 1
\label{45}
\end{eqnarray}
therefore the turbulent scattering of the ions will be dominated by the electrostatic turbulent electric field and the methodology developed in Ref.\cite{Mikhailenko-2011} is completely applicable here.
\begin{eqnarray}
&\displaystyle
\left[-\left(1+T \right)+A_{0i}\left(t\right) \right]\Phi\left(\mathbf{k},t\right)
\nonumber \\
&\displaystyle
+ i k_{y}v_{di}\int\limits^{t}_{t_{0}}dt_{1}
\Phi\left(\mathbf{k},t_{1}\right)A_{0i}\left(t, t_{1}\right)\exp \left(-t^{2}\int\limits^{t}_{t_{1}}d\hat{t}
\frac{C\left(\mathbf{k},\hat{t}\right)}{\hat{t}^{2}}\right)
\nonumber \\
&\displaystyle
+i\sqrt{\frac{\pi}{2}}\frac{k_{y}v_{de}}{k_{z}v_{Te}}\Phi\left(\mathbf{k},t\right)+T\sqrt{\frac{\pi}{2}}\frac{1}{k_{z}v_{Te}}\frac{\partial \Phi\left(\mathbf{k},t\right)}{\partial t}\nonumber \\
&\displaystyle +\frac{1}{k_{z}c}T\left( 1-\sqrt{\frac{\pi}{2}}\frac{1}{k_{z}v_{Te}}\frac{\partial}{\partial t}\right) \left(i\frac{\partial A\left(\mathbf{k},t\right)}{\partial t}-k_{y}v_{de} A\left(\mathbf{k},t\right) \right)=0, \label{46}
\end{eqnarray} 
where the approximation 
\begin{eqnarray}
&\displaystyle\left\langle\left(\mathbf{k}\left(t\right)
\delta\mathbf{r}\left(t\right)-\mathbf{k}\left(t_{1}\right)
\delta\mathbf{r}\left(t_{1}\right)\right)^{2}\right\rangle
\approx\frac{v_{\bot}^{2}}{2\omega^{2}_{ci}}\left\langle\left(\mathbf{k}\left(t\right)
\delta\phi\left(t\right)-\mathbf{k}\left(t_{1}\right)
\delta\phi\left(t_{1}\right)\right)^{2}\right\rangle \nonumber\\ &
\displaystyle \approx 2t^{2}\int\limits^{t}_{t_{1}}d\hat{t}
\frac{C\left(\mathbf{k},\hat{t}\right)}{\hat{t}^{2}}\label{47}
\end{eqnarray}
was used. In Eq.(\ref {47}) the function $C\left(\mathbf{k},\hat{t}\right)$ is determined iteratively by the equation
\begin{eqnarray}
&\displaystyle
C\left(\mathbf{k},\hat{t}\right)=\frac{c^{2}k^{2}_{y}\left(V'_{0}\hat{t}\right)^{6}v_{\bot}^{2}}
{8B^{2}\omega_{ci}^{2}}\int
d\mathbf{k}_{1}k^{4}_{1y}\left|\varphi\left(\mathbf{k}_{1},
\hat{t}\right)\right|^{2}\frac{C\left(\mathbf{k}_{1},\hat{t}\right)}
{\omega^{2}\left(\mathbf{k}_{1}\right)}\label{48}
\end{eqnarray}
with 
\begin{eqnarray}
&\displaystyle \left|\varphi\left(\mathbf{k},
\hat{t}\right)\right|^{2}=\left|\varphi\left(\mathbf{k},
\hat{t}_{0}\right)\right|^{2}\exp \left(\frac{1}{2}\sqrt{\frac{\pi}{3}}k_{y}v_{de}k_{y}\rho_
{i}\left(\frac{1}{\beta}\frac{T_{i}}{T_{e}}\frac{m_{e}}{m_{i}}\right)^{1/2}V'_{0}\hat{t}^{2}\right)
\label{49}
\end{eqnarray}
The integration in parts performed similar with Eq.(\ref {30}) gives
\begin{eqnarray}
&\displaystyle 
i k_{y}v_{di}\int\limits^{t}_{t_{0}}dt_{1}
\Phi\left(\mathbf{k},t_{1}\right)A_{0i}\left(t, t_{1}\right)\exp \left(-t^{2}\int\limits^{t}_{t_{1}}d\hat{t}
\frac{C\left(\mathbf{k},\hat{t}\right)}{\hat{t}^{2}}\right)\nonumber \\
&\displaystyle =-\frac{k_{y}v_{di}A_{0i}\left(t\right)}{\omega\left(\mathbf{k}, t\right)-iC\left(\mathbf{k}, t\right)}\Phi\left(\mathbf{k}\right)e^{-i\int\limits^{t}_{t_{0} }\omega\left(\mathbf{k}, t_{1} \right)dt_{1}}
+\frac{k_{y}v_{di}A_{0i}\left(t, t_{0}\right)}{\omega\left(\mathbf{k}, t_{0}\right)-iC\left(\mathbf{k}, t_{0}\right)}\Phi\left(\mathbf{k}\right)\nonumber \\
&\displaystyle +k_{y}v_{di}\Phi\left(\mathbf{k}\right)\int\limits^{t}_{t_{0}}dt_{1}\frac{i}{\left(\omega\left(\mathbf{k}, t_{1}\right)-iC\left(\mathbf{k}, t_{0}\right)\right)^{2}}
\nonumber \\
&\displaystyle \times\frac{d}{dt_{1}}\left(
e^{-i\int\limits^{t_{1}}_{t_{0} }\omega\left(\mathbf{k}, t_{2} \right)dt_{2}-t^{2}\int\limits^{t}_{t_{1}}d\hat{t}
\frac{C\left(\mathbf{k},\hat{t}\right)}{\hat{t}^{2}}}\right)\frac{d}{dt_{1}}\left(A_{0i}\left(t, t_{1}\right)\right)
\label{50}
\end{eqnarray}
Retaining in the right-hand side of Eq.(\ref {50}) only the first term, which is exponentially growing with time, we obtain the following renormalized form of Eq.(\ref {31}), in which the scattering of ions by the ensemble of the shearing drift-Alfven waves is accounted for:
\begin{eqnarray}
&\displaystyle \left[1+T-\left(1-\frac{k_{y}v_{di}}{\omega\left(\mathbf{k}, t\right)-iC\left(\mathbf{k}, t\right)} \right)A_{0i}\left(t \right)
+iT\sqrt{\frac{\pi}{2}}\frac{\left( \omega\left(\mathbf{k}, t \right)-k_{y}v_{de}\right) }{k_{z}v_{Te}} \right]\Phi\left(\mathbf{k}\right)\nonumber \\
&\displaystyle
-T\frac{\left( \omega\left(\mathbf{k}, t\right)-k_{y}v_{de}\right)}{k_{z}c}\left(1+i\sqrt{\frac{\pi}{2}}\frac{\omega\left(\mathbf{k}, t\right)}{k_{z}v_{Te}} \right)A\left(\mathbf{k}\right)=0 \label{51}
\end{eqnarray}
System (\ref {31}), (\ref {51}) gives the  equation for the function $\omega\left(\mathbf{k}, t\right)$. With approximation 
\begin{eqnarray}
&\displaystyle
\left(\omega\left(\mathbf{k}, t\right)-iC\left(\mathbf{k}, t\right)\right)^{-1}\approx\omega^{-1}\left(\mathbf{k}, t\right)+iC\left(\mathbf{k}, t\right)\omega^{-2}\left(\mathbf{k}, t\right)
\label{52}
\end{eqnarray}
that equation has a form
\begin{eqnarray}
&\displaystyle \left(\omega\left(\mathbf{k}, t\right)-k_{y}v_{de}\right)\left(\omega^{2}\left(\mathbf{k}, t\right)-\omega\left(\mathbf{k}, t\right)k_{y}v_{di}-k_{z}^{2}v^{2}_{A}\right)\nonumber \\
&\displaystyle
=k_{z}^{2}v^{2}_{A}\left[ k^{2}_{\bot}\left(t\right)\rho^{2}_{s}\left(\omega\left(\mathbf{k}, t\right)-k_{y}v_{di}\right)-iC\left(\mathbf{k}, t\right)\frac{k_{y}}{\omega\left(\mathbf{k}, t\right)}\right] \left(1-i\sqrt{\frac{\pi}{2}}\frac{\omega\left(\mathbf{k}, t\right)}
{k_{z}v_{Te}} \right). \label{53}
\end{eqnarray}
The solution for the nonlinear frequency mismatch, $\delta\omega\left(\mathbf{k}, t\right)$, follows from (\ref {53})
\begin{eqnarray}
&\displaystyle \delta\omega\left(\mathbf{k}, t\right)=\pm k_{z}v_{A}\left[k^{2}_{\bot}\left(t \right)\rho^{2}_{s}-i\frac{C\left(\mathbf{k}, t\right)}{\omega_{1}+|k_{y}v_{di}|}\right]^{1/2} \left(1-\frac{i}{2}\sqrt{\frac{\pi}{2}}\frac{k_{y}v_{de}}{k_{z}v_{Te}}\right). \label{54}
\end{eqnarray}
where $\omega_{1}$ is determined by Eq.(\ref{39}). Eq.(\ref {54}) displays the unexpected result: the turbulent scattering of the ions does not suppressed the drift-Alfven instability. Instead, that effect accelerates the  non-modal growth of the potentials with time. The quenching of the instability occurs due to violation of the  
frequency matching condition on time $t\gtrsim \left(V'_{0}k_{y}\rho_{i}\right)^{-1}$. 

\section*{CONCLUSIONS}
In this paper, we develop the non-modal theory of the kinetic drift-Alfven instability of plasma shear flow. In this theory, the shear flow reveals as the time-dependent effect of the finite Larmor radius in the system of integral equations for the electrostatic potential and for along-magnetic-field component of the vector potential. This effect is of principal importance for turbulence evolution in plasma shear flows\cite{Mikhailenko-2011}. It consists in the interaction of ions undergoing cyclotron motion with inhomogeneous electric field of sheared modes, which due to their distortion by shear flows have time dependent wave number in the laboratory frame of reference. In Ref.\cite{Mikhailenko-2011} we obtained, that this effect reveals in the non-modal decrease with time the frequency and the growth rate of the electrostatic kinetic drift instability by shear flow and results in the enhanced suppression of this instability. In this paper we find, however, that electromagnetic counterpart of that instability - kinetic drift-Alfven instability has principally different behaviour in shear flow. We derive, that due to shear flow the growth rate of that instability experiences the non-modal transient growth with time and, as a result, the anomalously fast development of the drift-Alfven instability occurs. The turbulent scattering of ions by the sheared drift-Alfven perturbations accelerates this growth. 

At time $t>\left(V'_{0}k_{y}\rho_{s}\right)^{-1}$, due to the non-modal growth of the mismatch of the drift and Alfven waves frequencies, resulted from time dependent finite ion Larmor radius effect, this instability transforms into electrostatic kinetic drift instability and separately damped Alfven waves. On that stage of the evolution, electrostatic drift instability experiences the linear and nonlinear damping and eventual suppression\cite{Mikhailenko-2011} conditioned by the velocity shear.

\begin{acknowledgments}
This work was funded by National R$\&$D Program through the National
Research Foundation of Korea(NRF) funded by the Ministry of Education, Science and
Technology (Grant No. 2012-M1A7A1A02-034918).
\end{acknowledgments}

\appendix
\section{THE DERIVATION OF THE FOURIER TRANSFORMED Eq.(7) BY USING LABORATORY FRAME SPATIAL COORDINATES.}
Here we display that  Eq.(\ref{7}) may be obtained without the transformation to the sheared 
coordinates (\ref{4}).  After the performing the transformation to convective coordinates (\ref{3}) 
in velocity space, Eq.(\ref{1}) becomes 
\begin{eqnarray}
&\displaystyle \frac{\partial f_{\alpha}}{\partial t}+\mathbf{v}\frac{\partial f_{\alpha}}
{\partial \mathbf{\hat{r}}} +V'_{0}\hat{x}\frac{\partial
f_{\alpha}}{\partial \hat{y}} + \omega_{c\alpha} v_{ y}
\frac{\partial f_{\alpha}}{\partial v_{\alpha
x}}-\left(\omega_{c\alpha}+V'_{0} \right) v_{\alpha
x}\frac{\partial f_{\alpha}}{\partial v_{\alpha y}} \nonumber
\\  &\displaystyle
=\frac{e_{\alpha}}{m_{\alpha}}\nabla \varphi \left(\mathbf{\hat{r}},t \right)\frac{\partial F_{0\alpha}}
{\partial \mathbf{\hat{v}}} -\frac{e_{\alpha}}{m_{\alpha}}\frac{v_{\alpha
z}}{c}\frac{\partial A_{z}}{\partial
\hat{x}}\frac{\partial F_{0\alpha}}{\partial v_{\alpha
x}}-\frac{e_{\alpha}}{m_{\alpha}}\frac{v_{\alpha
z}}{c}\frac{\partial A_{z}}{\partial \hat{y}}\frac{\partial F_{0\alpha}}{\partial v_{\alpha
y}}\nonumber
\\  &\displaystyle +\frac{1}{c}\frac{e_{\alpha}}{m_{\alpha}}\left[\frac{\partial A_{z}}{\partial t}+V'_{0}\hat{x}\frac{\partial A_{z}}{\partial \hat{y}} + v_{\alpha x}
\frac{\partial A_{z}}{\partial
\hat{x}} +v_{\alpha y}\frac{\partial A_{z}}{\partial \hat{y}} \right]\frac{\partial F_{0\alpha}}{\partial v_{\alpha
z}}.\label{A1}
\end{eqnarray}
After the  Fourier transforming of Eq.(A1) over the coordinates $\hat{x}, \hat{y}, \hat{z}$ in laboratory frame as
\begin{eqnarray}
&\displaystyle f_{\alpha} \left(\mathbf{v}, \mathbf{\hat{k}}, t\right) 
= \int f_{\alpha} \left(\mathbf{v}, \mathbf{\hat{r}},
t\right)e^{-i\mathbf{\hat{k}}\mathbf{\hat{r}}} d\mathbf{\hat{k}}\label{A2}
\end{eqnarray} 
we receive the following equation for $f_{\alpha}\left(t, \mathbf{\hat{k}}, \mathbf{v} \right)$:

\begin{eqnarray}
&\displaystyle \frac{\partial f_{\alpha}}{\partial
t}-V'_{0}\hat{k_{y}}\frac{\partial f_{\alpha}}{\partial
\hat{k_{x}}}+i\left( \hat{\mathbf{k}}
\hat{\mathbf{v}}\right)f_{\alpha}+\omega_{c\alpha} v_{\alpha y}
\frac{\partial f_{\alpha}}{\partial v_{\alpha
x}}-\left(\omega_{c\alpha}+V'_{0} \right) v_{\alpha
x}\frac{\partial f_{\alpha}}{\partial v_{\alpha y}} \nonumber  \\
&\displaystyle =i\frac{e_{\alpha}}{m_{\alpha}}\hat{\mathbf{k}}
\varphi\left(\hat{\mathbf{k}},t\right)\frac{\partial
F_{0\alpha}}{\partial\mathbf{v}}-i\frac{e_{\alpha}}{m_{\alpha}c}\hat{k}_{x}v_{z}
A_{z}\left(\mathbf{\hat{k}}, t\right)\frac{\partial F_{0\alpha}}{\partial v_{x}}
-i\frac{e_{\alpha}}{m_{\alpha}c}\hat{k}_{y}v_{z}
A_{z}\left(\mathbf{\hat{k}}, t\right)\frac{\partial F_{0\alpha}}{\partial v_{y}} \nonumber  \\
&\displaystyle +i\frac{e_{\alpha}}{m_{\alpha}c}\left(\hat{k}_{y}v_{x}+\hat{k}_{y}v_{y}\right)
A_{z}\left(\mathbf{\hat{k}}, t\right)\frac{\partial F_{0\alpha}}{\partial v_{z}}\nonumber  \\
&\displaystyle +\frac{e_{\alpha}}{m_{\alpha}c}\left(\frac{\partial A_{z}}{\partial t}-V'_{0}\hat{k}_{y}\frac{\partial A_{z}}{\partial \hat{k}_{x}} \right)\frac{\partial F_{0\alpha}}{\partial v_{z}}  .\label{A3} 
\end{eqnarray}
For the deriving from Eq.(\ref{A3}) the equation which 
couples $f_{\alpha}$ with potentials $\varphi$ and $A_{z}$ of the separate spatial Fourier mode, 
as it is in Eq.(\ref{7}), we have to exclude from Eq.(\ref{A3}) the differential operator $-V'_{0}\hat{k}_{y}\frac{\partial f_{\alpha}}{\partial \hat{k}_{x}}$, due to which the 
Fourier mode of $f_{\alpha}$ appears to be coupled with all Fourier modes of the 
electrostatic and vector potentials and depends on the integral of $\varphi$ and $A_{z}$ over wave-number space.
The characteristic equation 
\begin{eqnarray}
&\displaystyle
dt=-\frac{d\hat{k}_{x}}{V'_{0}\hat{k}_{y}}
\label{A4} 
\end{eqnarray}
gives the solution $\hat{k}_{x}+V'_{0}t\hat{k}_{y}=K_{x}$, where $K_{x}$ 
as the integral of Eq.(\ref{10}) is time independent. It reveals that $f_{\alpha}
=f_{\alpha}\left(K_{x}, \hat{k}_{y}, \hat{k}_{z}, t\right)=f_{\alpha}\left(\hat{k}_{x}
+V'_{0}t\hat{k}_{y}, \hat{k}_{y}, \hat{k}_{z}, t\right)$, 
i.e. the wave number components $\hat{k}_{x}$ and $\hat{k}_{y}$ have to be changed 
in such a way that $\hat{k}_{x}+V'_{0}t\hat{k}_{y}$ leaves unchanged with time. 
If we use  $\hat{k}_{x}=K_{x}-V'_{0}t\hat{k}_{y}$ in Eqs.(\ref{A2}) 
and (\ref{A3}), we obtain for the electrostatic potential the presentation (\ref{8}), and  
we obtain Eq.(\ref{7}) for $f_{\alpha}$, with time independent 
$K_{x}=k_{x}$, $\hat{k}_{y}=k_{y}$, $\hat{k}_{z}=k_{z}$. Note, that the separate Fourier harmonic in $\hat{\mathbf{k}}$ space may be extracted, if potentials $\varphi$ and $A_{z}$ also are 
determined as the shearing modes, i.e. $\varphi=\varphi\left(\hat{k}_{x}+V'_{0}t\hat{k}_{y},  
\hat{k}_{y}, \hat{k}_{z}, t \right)=\varphi\left(k_{x},  
k_{y}, k_{z}, t \right)$. The obtained results 
prove, that the solution of the Vlasov equation for times $t\geqslant \left( V'_{0}\right)^{-1}$ 
in the form of the separate Fourier harmonic may be obtained only in 
convected-sheared coordinates. That solution reveals in the laboratory frame 
as a shearing mode (\ref{8}) with time dependent $x$-component of the  wave number.

\end{document}